\documentclass[%
preprint,
superscriptaddress,
amsmath,amssymb,
]{revtex4-1}
\usepackage{color}
\usepackage{graphicx}
\usepackage{subfigure}
\usepackage{mathtools}
\usepackage{amsmath}
\usepackage{dcolumn}
\usepackage{bm}

\begin{document}


\title{Broadband Tunable Phase Shifter For Microwaves}

\author{Jinli Zhang}
\affiliation{College of Electronic Information and Optical Engineering, Nankai University, Tianjin, 300071, China}
\affiliation{QCD Labs, QTF Centre of Excellence, Department of Applied Physics, Aalto University, P.O. Box 13500, FI-00076 Aalto, Finland}

\author{Tianyi Li}
\affiliation{QCD Labs, QTF Centre of Excellence, Department of Applied Physics, Aalto University, P.O. Box 13500, FI-00076 Aalto, Finland}
\author{Roope Kokkoniemi}
\affiliation{QCD Labs, QTF Centre of Excellence, Department of Applied Physics, Aalto University, P.O. Box 13500, FI-00076 Aalto, Finland}
\author{Chengyu Yan}
\affiliation{QCD Labs, QTF Centre of Excellence, Department of Applied Physics, Aalto University, P.O. Box 13500, FI-00076 Aalto, Finland}
\author{Wei Liu}
\affiliation{QCD Labs, QTF Centre of Excellence, Department of Applied Physics, Aalto University, P.O. Box 13500, FI-00076 Aalto, Finland}
\author{Matti Partanen}
\affiliation{QCD Labs, QTF Centre of Excellence, Department of Applied Physics, Aalto University, P.O. Box 13500, FI-00076 Aalto, Finland}
\author{Kuan Yen Tan}
\affiliation{QCD Labs, QTF Centre of Excellence, Department of Applied Physics, Aalto University, P.O. Box 13500, FI-00076 Aalto, Finland}
\author{Ming He}
\affiliation{College of Electronic Information and Optical Engineering, Nankai University, Tianjin, 300071, China}
\author{Lu Ji}
\affiliation{College of Electronic Information and Optical Engineering, Nankai University, Tianjin, 300071, China}
\author{Leif Gr\"onberg}
\affiliation{VTT Technical Research Centre of Finland Ltd, P.O. Box 1000, FI-02044 VTT, Finland}
\author{Mikko M\"ott\"onen}
\affiliation{QCD Labs, QTF Centre of Excellence, Department of Applied Physics, Aalto University, P.O. Box 13500, FI-00076 Aalto, Finland}
\affiliation{VTT Technical Research Centre of Finland Ltd, P.O. Box 1000, FI-02044 VTT, Finland}
\email{mikko.mottonnen@aalto.fi}




\date{\today}

\begin{abstract}
We implement a broadly
tunable phase shifter for microwaves based on superconducting quantum interference devices (SQUIDs) and study it both  experimentally and theoretically. At  different frequencies, a unit transmission coefficient, $|S_{21}|=1$, can be theoretically achieved along a curve where the phase shift is controllable by magnetic flux. The fabricated device consists of three equidistant SQUIDs interrupting a transmission line. We model each SQUID embedded at different positions along the transmission line with two parameters, capacitance and inductance, the values of which we extract from the experiments. In our experiments, the tunability of the phase shift varies from from $0.07\times\pi$ to $0.14\times\pi$ radians along the full-transmission curve with the input frequency ranging from 6.00 to 6.28~GHz. The reported measurements are in good agreement with simulations, which is promising for future design work of phase shifters for different applications.
\end{abstract}

\maketitle

\section{introduction}
Recent progress in superconducting microwave electronics has inspired research on a more complete toolbox for quantum engineering~\cite{krantz2019quantum}. Here, superconducting circuits with Josephson junctions 
exhibit a solid and scalable technology platform stemming from their mature lithographic fabrication processes~\cite{devoret2013superconducting}.
During the recent decades, fascinating superconducting rf components for the toolbox have been demonstrated, such as Josephson parametric amplifiers~\cite{yamamoto2008flux,macklin2015near}, kinetic inductance travelling-wave amplifiers~\cite{eom2012wideband,vissers2016low}, switches~\cite{naaman2016chip,pechal2016superconducting}, circulators~\cite{muller2018passive,chapman2019design}, isolators~\cite{abdo2019active}, beam splitters~\cite{hoffmann2010superconducting,mariantoni2010planck}, phase shifters~\cite{kokkoniemi2017flux,naaman2017josephson}, and photon detectors~\cite{govenius2016detection,besse2018single,kono2018quantum}.
In the future, these may be integrated into monolithic circuits for sophisticated quantum signal processing.

To further improve the ability to process quantum microwave information, a quickly tunable, compact, and lossless phase shifter for microwave photons operating over a broad frequency band is a highly desirable tool, not only to tailor propagating single-photon states~\cite{Eichler_PRL_tomography} but also to tune the phase of on-chip coherent microwave sources~\cite{cassidy2017demonstration}. If such sources are further augmented with quantum-circuit refrigerators~\cite{tan2017quantum,Silveri19}, bulky room temperature signal generators could be replaced by devices on a single chip. Such tool would be highly desirable for scaling up a quantum computer~\cite{arute2019quantum}.

Interestingly, transfer of quantum states between distant stationary qubits has been achieved utilizing propagating microwave photons~\cite{kurpiers2018deterministic}. Such photons are also required for far-field microwave quantum communication. Utilization of a tunable phase shifter in such schemes provides opportunities for detailed control of the quantum states of the propagating photons. For example, the phase shifter would allow for the preparation of an arbitrary squeezing angle of squeezed states for secure communication~\cite{pogorzalek2019secure}.

In this letter, we experimentally realize a tunable phase shifter based on three equidistant SQUIDs in a transmission line.
We improve upon our previous design~\cite{kokkoniemi2017flux} by adopting differential flux bias lines instead of single-ended flux bias lines to greatly decrease the cross coupling of the SQUID fluxes, from which the previous design suffered. Consequently, we demonstrate that we can tune the operating frequency of the device by 280~MHz.
We utilize our theoretical model complemented by numerical computations to control the phase shifter such that it provides essentially unit transmission throughout this frequency range of interest. Thus our work is a significant step towards an extended toolbox of superconducting microwave components.
\begin{figure}[htb]
\includegraphics[width=8.5 cm]{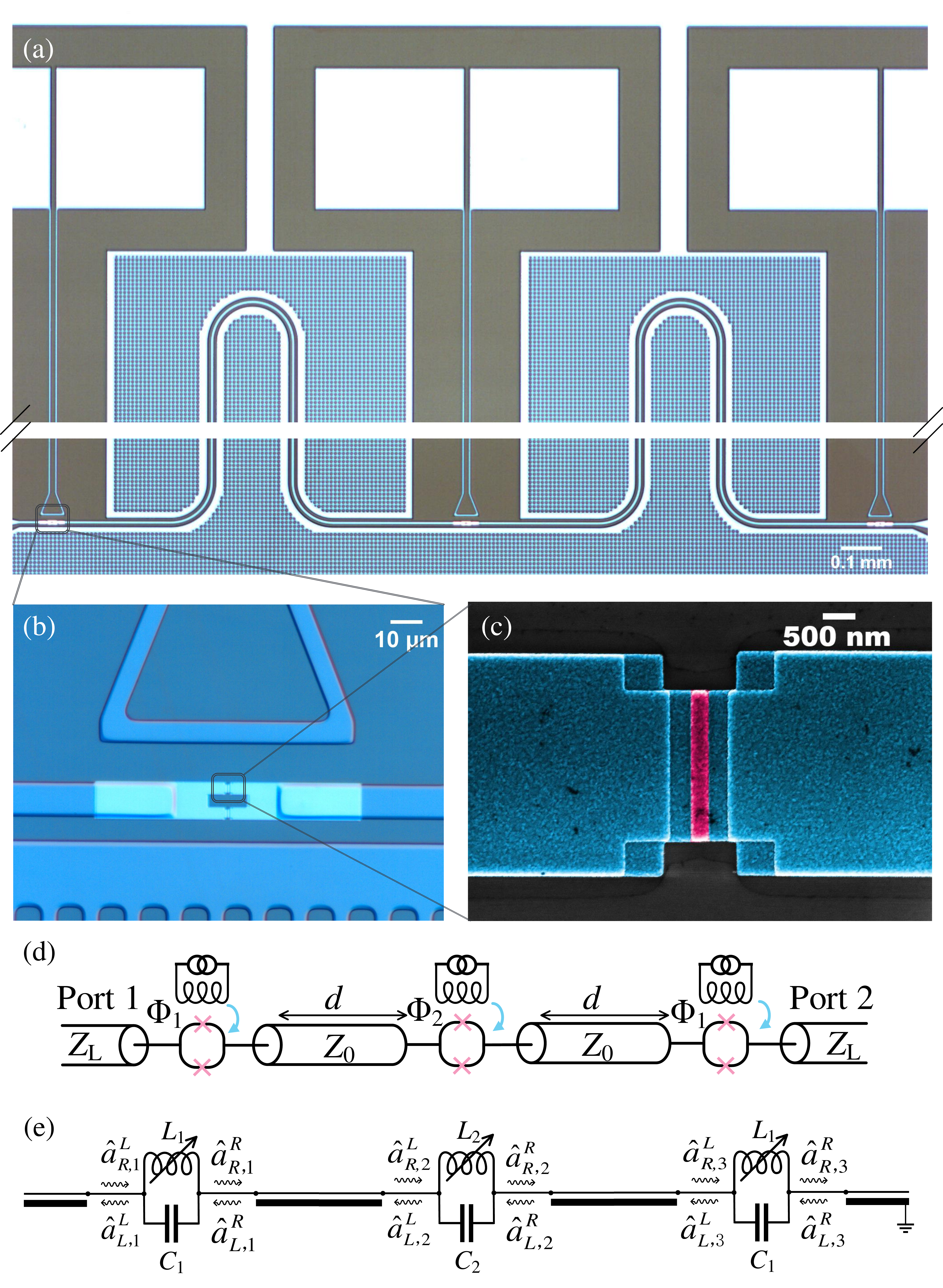}
\caption{\label{sample} 
Colored optical image of the measured device showing (a) three SQUIDs equidistantly placed on a coplanar-waveguide transmission line (CPWTL) and (b)  a single SQUID next to its differential magnetic-flux bias line. (c) Colored scanning-electron-microscope image of a single Josephson junction (red color) of a SQUID. (d) Schematic diagram of the phase shifter. Three SQUIDs are galvanically connected to a CPWTL with a distance $d = 6.41$~mm. The characteristic impedance of the CPWTL and of the two ports are both $Z_\textrm{L} = Z_{0} = 50$~$\Omega$. Three differential flux bias lines are used to tune the magnetic fluxes through the SQUIDs. (e) As (d) but the SQUIDs are approximated by tunable lumped-element $LC$ resonators. Operators $\hat{a}^L_{Lk}$ and $\hat{a}^R_{Lk}$ ($k=1,2,3$) denote annihilation operators of the left-moving wave on the left and right side of the resonator $k$, respectively, and $\hat{a}^L_{Rk}$ and $\hat{a}^R_{Rk}$ denote the corresponding right-moving operators.}
\end{figure}

\section{Theoretical model}
The considered phase shifter is composed of three SQUIDs connected by two coplanar-waveguide transmission lines (CPWTLs) of equal length $d = 6.41$~mm as shown in Fig.~\ref{sample}(a). The use of CPWTLs provides a possibility for achieving a broad bandwidth and unit transmission in a finite frequency range in contrast to a phase shifter where microwaves reflect from a resonator~\cite{naaman2017josephson}. 
In our theoretical model, each SQUID is treated as a parallel connection of a tunable inductor and a capacitor, thus forming an $LC$ oscillator. The inductance of an ideal SQUID, $L_k$, can be modulated by applying external magnetic flux as $L_k=\Phi_0/(4 \pi I_c|\cos(\pi\Phi_k/\Phi_0)|)$, where $\Phi_0$ is the magnetic flux quantum, $I_c$ is the critical current of the SQUID and $\Phi_k$ ($k=1,2$) is the external magnetic-flux threading the loop of SQUID $k$. In contrast to Ref.~\cite{kokkoniemi2017flux}, we have not assumed all SQUIDs to be identical, but allow for the center SQUID in our model to have a different capacitance, $C_2$, than that of the side SQUIDS, $C_1$.

Let us consider quantum scattering of microwaves from the leftmost $LC$ oscillator in Fig.~\ref{sample}(e). In the Heisenberg picture, the quantum network theory~\cite{yurke1984quantum} yields
\begin{equation}\label{mainequation}
\splitfrac{Z_0 (\hat{a}^L_{L1}+\hat{a}^L_{R1}-\hat{a}^R_{L1}-\hat{a}^R_{R1})=i L_1 \omega  [\hat{a}^L_{L1}-\hat{a}^L_{R1}}{-i C_1 \omega  Z_0
   (\hat{a}^L_{L1}+\hat{a}^L_{R1}-\hat{a}^R_{L1}-\hat{a}^R_{R1})]},
\end{equation}
where $\hat{a}^L_{L1}$ and $\hat{a}^R_{L1}$ denote annihilation operators of the left-moving wave on the left and on the right side of the left resonator, respectively, and $\hat{a}^L_{R1}$ and $\hat{a}^R_{R1}$ denote the corresponding right-moving operators. Considering Kirchhoff’s current law and Fourier expansion of the charge operators with annihilation operators of signal quanta, one can writes
\begin{equation}\label{boundary1}
\hat{a}^L_{R1}-\hat{a}^L_{L1}=\hat{a}^R_{R1}-\hat{a}^R_{L1}.
\end{equation}
Similar equations and boundary conditions can be obtained by analyzing the middle and right oscillators in Fig.~\ref{sample}(e).
The CPWTLs between the three oscillators generate a delay which converts into a phase change of the propagating signal, $\phi = \omega d/v$, where $\omega$ is the angular frequency of the microwave radiation and $v$ is its speed. This can be written as
\begin{equation}\label{boundary2}
\begin{aligned}
\hat{a}^R_{L1}=\hat{a}^L_{L2} e^{i \phi },\\
\hat{a}^R_{R1}=\hat{a}^L_{R2} e^{i \phi },\\
\hat{a}^R_{L2}=\hat{a}^L_{L3} e^{i \phi },\\
\hat{a}^R_{R2}=\hat{a}^L_{R3} e^{i \phi }.
\end{aligned}
\end{equation}
We solve Eq.~(\ref{mainequation}) utilizing the boundary conditions Eq.~(\ref{boundary1}) and Eq.~(\ref{boundary2}) to obtain the transmission coefficient
\begin{widetext}
\begin{equation}\label{Tcoef}
S_{21} = \frac{8 Z_0^3 e^{2 i \phi } A^2 B[2Z_0 A-i L_1 \omega  \left(-1+e^{2 i \phi }\right)]^{-1}}
{4 Z_0^2A B+2 i \omega  Z_0 \left\{-L_2+L_1 \left[C_1 L_2 \omega ^2+C_2 L_2 \omega ^2+e^{2 i \phi }B-1\right]\right\}+L_1 L_2 \omega ^2 \left(-1+e^{2 i \phi }\right)},
\end{equation}
\end{widetext}
where
\begin{equation}
A = C_1 L_1 \omega ^2-1,
\end{equation}
\begin{equation}
B = C_2 L_2 \omega ^2-1.
\end{equation}
Assuming that the SQUID inductances are arbitrarily tunable, we may choose

\begin{equation}\label{inductance}
\begin{aligned}
L_1 = \frac{2 Z_0 \sin \left(\frac{\theta }{2}\right)}{
\omega  \{2 C_1 \omega  Z_0 \sin
   \left(\frac{\theta }{2}\right)
   -\cos \left[\frac{1}{2} (\theta +4 \phi )\right]+\cos \left(\frac{\theta
   }{2}\right)\}}
   ,\\
L_2 = \frac{4 Z_0 \sin \left(\frac{\theta }{2}\right) \cos \left[\frac{1}{2} (\theta +4 \phi )\right]}{
\omega  [2
   C_2 \omega  Z_0 \sin (\theta +2 \phi )
   -2 C_2 \omega  Z_0 \sin (2 \phi )+\cos (2 \phi )-1]}
   ,
\end{aligned}
\end{equation}
where $\theta$ is a free real-valued parameter fixing our choice of the flux points. There is only a single free parameter since we have chosen the parametrization of the inductance such that microwave reflections from the circuit vanish.
Insertion of Eq.~(\ref{inductance}) into Eq.~\eqref{Tcoef} yields $S_{21} = e^{i(\theta+2\phi)}$. Thus the device exhibits full transmission and a phase shift of $\theta$ in addition to $2\phi$ arising from the transmission line of length $2d$. The tunability range of $\theta$ depends on the ability to implement the inductances according to Eq.~(\ref{inductance}). In particular, vanishing or negative inductances are not feasible in the implementation described in Fig.~\ref{sample}.


\begin{figure*}[hbt]
\includegraphics[width=1\linewidth]{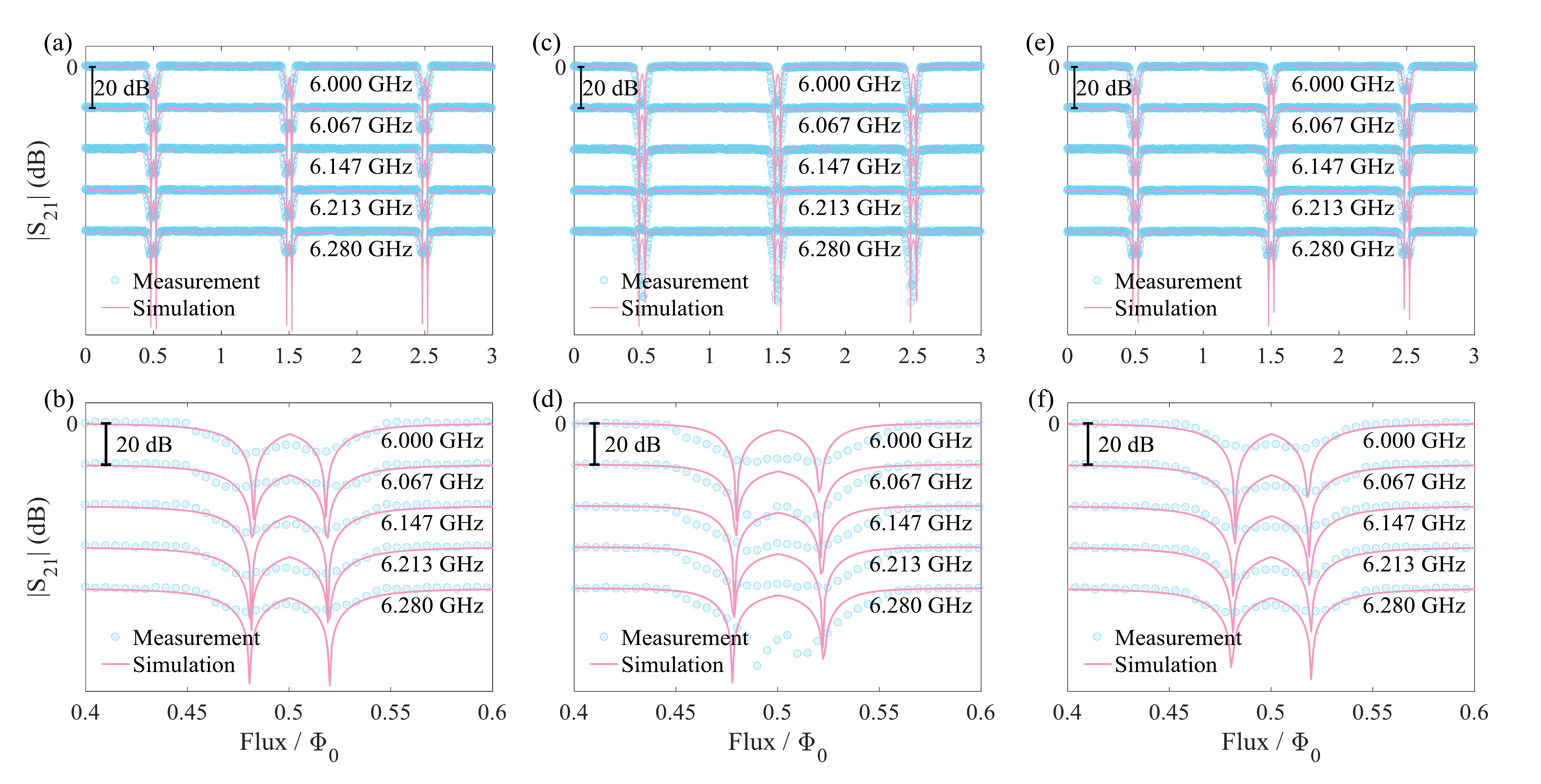}
\caption{\label{single-SQUID} 
Magnitude of the transmission coefficient $|S_{21}|$ as a function of the (a), (b) left, (c), (d) middle, and (e), (f) right SQUID flux at the indicated frequencies in the measurement (blue circles) and classical-circuit simulation (solid lines).
The fluxes which are not swept are held close to zero. 
The curves are vertically offset by 20 dB for the sake of clarity. 
In the simulation, the left and right SQUIDs are identical having a capacitance of 180 fF and a critical current $I_{c}=1.5$~$\mu $A. The capacitance and $I_{c}$ of the middle SQUID are 170 fF and 1.24~$\mu$A, respectively.}
\end{figure*}

\section{experimental results}
To implement the above theoretical scheme, we fabricate a sample adopting shadow evaporation and load it into a dilution refrigerator operating at 15 mK. On-chip differential bias lines are utilized to produce a bias magnetic field at each SQUID with low crosstalk. The device is reciprocal and symmetric with respect to the left and right SQUIDs, which renders it convenient to integrate the phase shifter with other on-chip components for future applications.

The power level of the probe signal at the device is kept below $-90$~dBm in order to keep the SQUIDs in the linear regime. Note that this is well above the single-photon level. Details of the measurement setup are given in Supplementary Materials.

\subsection{Characterization}
We begin the characterization of our device by first focusing on a single SQUID at a time. Namely, we measure the transmission coefficient of the device as a function of the magnetic-flux bias of each SQUID at a time, ideally leaving the other two SQUIDs at a constant magnetic field.  Figure~\ref{single-SQUID}(a) shows the transmission amplitude $|S_{21}|$ as a function of the flux bias of the left SQUID for three periods at five different frequencies ranging from 6.00 GHz to 6.28 GHz. The amplitude changes very gently around integer flux quanta but exhibits sharp drops in the vicinity of half-integer flux values where the SQUID inductance ideally diverges. In fact, we observe two drops in the transmission amplitude around each half-integer flux points since the parallel $LC$ oscillator achieves its maximum impedance at finite inductance where the $LC$ resonance matches with the probe frequency.

Figure~\ref{single-SQUID}(b) shows the transmission amplitude from $0.4\times\Phi_0$ to $0.6\times\Phi_0$ to demonstrate that the experimental results are in good agreement with simulations where we use the SQUID capacitance and the critical current $I_c$ as fitting parameters. 
Consequently, we obtain 1.5~$\mu$A for the critical current of the left SQUID, which yields approximately 0.44 nH of inductance. Thus at zero flux, the magnitude of the transmission coefficient is close to unity since the impedance of a zero-flux SQUID is given by $1/[(1/i\omega L_1)+i\omega C_1] = i18.7 $~$\Omega$ for the obtained capacitance $C_1=180$~fF at the frequencies of interest. In Figs.~\ref{single-SQUID}(c)--\ref{single-SQUID}(f), we apply this method to the middle and right SQUIDs and obtaine similar results as for the left SQUID. In our model, we set the critical current and capacitance of the right SQUID equal to those of the left SQUID, 1.5~$\mu$A and 180~fF, respectively. The critical current of the middle SQUID is 1.24~$\mu$A and capacitance is 170~fF. 

\begin{figure*}[!b]
\includegraphics[width=1\linewidth]{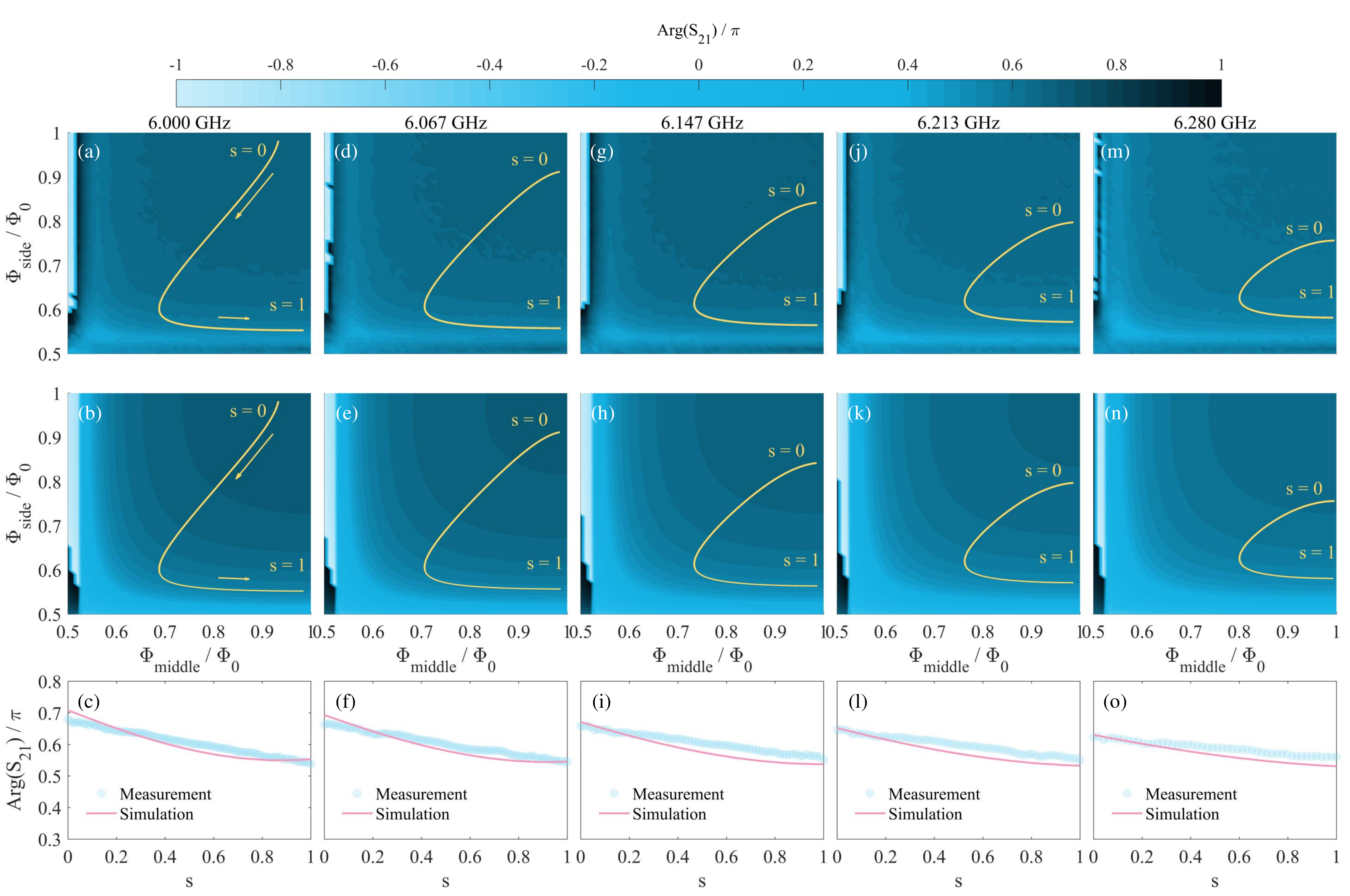}
\caption{\label{tri:phase}
(a), (d), (g), (j), (m) Measured phase of the transmission coefficient as a function of the side and middle SQUID magnetic fluxes for a half of a period at the indicated frequencies. (b), (e), (h), (k), (n) As above but for the simulated phase shift. The yellow line denotes the curve along which the magnitude of the transmission coefficient is unity in the simulations. The parameter $s$ defines the coordinates on the curve as indicated. (c), (f), (i), (l), (o) The measured (blue circles) and simulated (solid line) phase of the transmission coefficient along the full-transmission curve as a function of the parameter $s$.
The simulation parameters are chosen as in Fig.~\ref{single-SQUID}.}
\end{figure*}
\begin{figure}[h]
\includegraphics[width=8.5 cm]{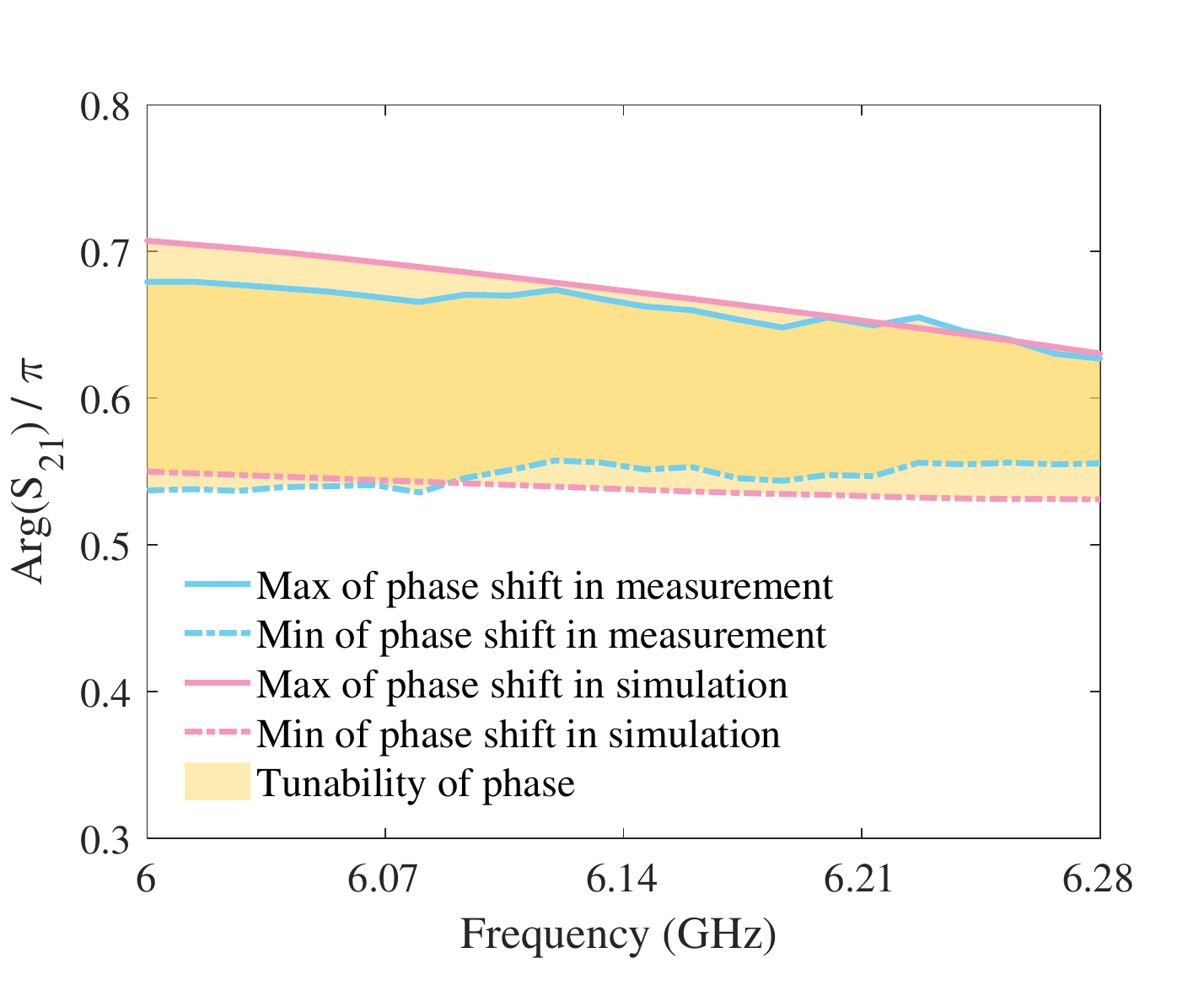}
\caption{\label{comparision} Maximum (solid lines) and minimum (dash-dotted lines) phase shift obtained in experiments (blue color) and in simulations (red color) corresponding to Fig.~\ref{tri:phase} as functions of frequency. The shading defines the region of achievable phase shifts, i.e., the tunablity region of the phase shifter.}
\end{figure}

\subsection{Phase shift and its tunability}
In Ref.~\cite{kokkoniemi2017flux}, this type of a phase shifter was challenging to operate at different frequencies because of inductive crosstalk. The applied bias current induced unwanted currents to the ground ground plane near all SQUIDs. To eliminate this effect, we redesigned the flux bias lines to be differential and removed the ground plane from its vicinity.

Consequently, we show in Fig.~\ref{tri:phase} the phase of the scattering parameter as a function of the magnetic flux threading the middle SQUID and both side SQUIDs at five different frequencies which match those of Fig.~\ref{single-SQUID}. The fluxes through the side SQUIDs are tuned to keep their inductances equal with each other.
Using bilinear interpolation in the flux plane, we extract the transmission coefficient along the full-transmission curve that follows Eqs.~(\ref{inductance}) for both the magnitude $|S_{21}|$ (see supplemental materials) and the phase Arg$(S_{21})$.
Figures~\ref{tri:phase}(c), \ref{tri:phase}(f), \ref{tri:phase}(i), \ref{tri:phase}(l), and \ref{tri:phase}(o) show the measured and simulated phase along the curve at five frequencies. Clear modulation of the phase is observable along the curve which indicates that we may tune the phase at will while keeping the transmission through the phase shifter close to unity.

Figure~\ref{comparision} summarizes the tunability of the phase shift in a dense frequency grid from~6.00 to 6.28~GHz. Here, we define the tunability as $\textrm{max}_s\textrm{Arg}(S_{21})- \textrm{min}_s\textrm{Arg}(S_{21})$ along the full-transmission curve parametrized by the parameter $s$ defined in Fig.~\ref{tri:phase}.  We observe that in the whole frequency range considered, the tunability is over $0.07\times\pi$ radians. The largest tunability is $0.14\times\pi$ radians at 6.013 GHz.
The tunability can be further optimized by fine-tuning the parameters of the transmission lines and of the SQUIDs in the design and fabrication steps. Such parameters include the sizes of the junctions which determine the critical currents and the capacitances of the junctions. Theoretically, an ideal three-SQUID device may achieve a maximum phase shift tunability approaching~$2\pi$ at the optimal frequency~\cite{kokkoniemi2017flux}.

At a given flux point on the curve of unit transmission, the transmission amplitude is relatively close to unity in the full studied frequency band from 4~GHz to 8~GHz, whereas the phase shift resulting from the phase shifter changes linearly by the amount of $\pi$ (see Supplementary Materials for data). This results in a phase error of roughly $10^{-3}\textrm{ rad}/\textrm{MHz}$. Thus also in a given flux point, the phase shifter works accurately in a relatively broad frequency band.
\section{conclusion}

In conclusion, we implemented a phase shifter composed of three equidistant SQUIDs in a transmission line. We presented an extension to the theory of the phase shifter by allowing the parameters of the middle SQUID to be different from those of the identical side SQUIDs. The undesired coupling from each flux bias line to the two distant SQUIDs was reduced by an improved design.
Consequently, by tuning the magnetic fluxes through the SQUIDs, we managed to observe significant phase shifts throughout a 280-MHz bandwidth from 6 GHz to 6.28 GHz. The experiments were found to be in good agreement with classical-circuit simulations, which provided us with estimates of the parameters of the SQUIDs. 

This tunable phase shifter exhibits potential for applications in quantum microwave signal generation and processing. In the future, we aim to optimize the phase shifter for operation in a broader frequency range and for a lager tunability of the phase shift. In addition, the phase shifter can be integrated with other microwave components, such as a quantum-circuit refrigerator~\cite{tan2017quantum,Silveri19} and a  microwave source~\cite{cassidy2017demonstration} to achieve a tunable single-chip source. Thus, this work paves the way for advanced cryogenic microwave devices and expands the quantum-engineering toolbox~\cite{krantz2019quantum}.
\begin{acknowledgments}
This research was financially supported by the European Research Council under grant no.~681311 (QUESS); by the European Commission through H2020 program project  QMiCS (grant agreement 820505, Quantum Flagship); by the Academy of Finland under its Centres of Excellence Program grant nos.~312300 and 312059 and grant nos.~308161, 314302, 316551, 318937, and 319579; and
 by the Alfred Kordelin Foundation, the Emil Aaltonen Foundation, the Vilho, Yrj\"o and Kalle V\"ais\"al\"a Foundation, the Jane and Aatos Erkko Foundation and the Technology Industries of Finland Centennial Foundation. We thank Jan Goetz and Juha Hassel for useful discussions and the provision of facilities and technical support by Aalto University at OtaNano~\---~Micronova Nanofabrication Centre.
\end{acknowledgments}

\bibliography{ZJL}

\end{document}